\documentclass[twocolumn]{aastex63}
\usepackage{lineno}
\usepackage{amsmath}
\usepackage{hyperref}
\usepackage{soul}

\received{2022 May 22}
\revised{2022 June 16}
\accepted{2022 June 21}

\begin{document}

\title{Fermi-LAT Detection of a GeV Afterglow from a Compact Stellar Merger}
\author[0000-0001-6863-5369]{Hai-Ming Zhang}
\affil{School of Astronomy and Space Science, Nanjing University, Nanjing 210023, China; xywang@nju.edu.cn}
\affil{Key laboratory of Modern Astronomy and Astrophysics (Nanjing University), Ministry of Education, Nanjing 210023, China}
\author[0000-0002-6036-985X]{Yi-Yun Huang}
\affil{School of Astronomy and Space Science, Nanjing University, Nanjing 210023, China; xywang@nju.edu.cn}
\affil{Key laboratory of Modern Astronomy and Astrophysics (Nanjing University), Ministry of Education, Nanjing 210023, China}
\author[0000-0001-5751-633X]{Jian-He Zheng}
\affil{School of Astronomy and Space Science, Nanjing University, Nanjing 210023, China; xywang@nju.edu.cn}
\affil{Key laboratory of Modern Astronomy and Astrophysics (Nanjing University), Ministry of Education, Nanjing 210023, China}
\author[0000-0003-1576-0961]{Ruo-Yu Liu}
\affil{School of Astronomy and Space Science, Nanjing University, Nanjing 210023, China; xywang@nju.edu.cn}
\affil{Key laboratory of Modern Astronomy and Astrophysics (Nanjing University), Ministry of Education, Nanjing 210023, China}
\author[0000-0002-5881-335X]{Xiang-Yu Wang}
\affil{School of Astronomy and Space Science, Nanjing University, Nanjing 210023, China; xywang@nju.edu.cn}
\affil{Key laboratory of Modern Astronomy and Astrophysics (Nanjing University), Ministry of Education, Nanjing 210023, China}

\begin{abstract}

It is usually thought that long-duration gamma-ray bursts (GRBs) are associated with massive star core collapse, whereas short-duration GRBs are associated with mergers of compact stellar binaries. The discovery of a kilonova associated with a nearby (350 Mpc) long-duration GRB—GRB 211211A, however, indicates that the progenitor of this long-duration GRB is a compact object merger. Here we report the Fermi-LAT detection of gamma-ray ($>100 {\rm \ MeV}$) afterglow emission from GRB 211211A, which lasts $\sim$20,000 s after the burst, the longest event for conventional short-duration GRBs ever detected. We suggest that this gamma-ray emission results from afterglow synchrotron emission. The soft spectrum of GeV emission may arise from a limited maximum synchrotron energy of only a few hundreds of MeV at $\sim$20,000 s. The unusually long duration of the GeV emission could be due to the proximity of this GRB and the long deceleration time of the GRB jet that is expanding in a low-density circumburst medium, consistent with the compact stellar merger scenario.
\end{abstract}


\section{Introduction}           
\label{sect:intro}
Gamma-ray bursts (GRBs) are usually divided into two populations \citep{Norris1984Natur.308..434N,Kouveliotou1993ApJ...413L.101K}: long
GRBs that originate from the core collapse of massive stars \citep{Galama1998Natur.395..670G} and
short GRBs formed in the merger of two compact objects \citep{Abbott2017ApJ848L12A}. While
it is common to divide the two populations at a duration of 2\,s for the prompt keV/MeV emission, classification based on duration only does not always correctly point
to the progenitor. Growing observations \citep{Ahumada2021NatAs...5..917A,GalYam2006Natur.444.1053G,Gehrels2006Natur.444.1044G,Zhang2021NatAs...5..911Z} have shown that multiple criteria (such as supernova/kilonova associations and host galaxy
properties) rather than burst duration only are needed to classify GRBs physically. 

GRB 211211A triggered the Burst Alert Telescope (BAT; \cite{2005SSRv..120..143B}) on board the Neil Gehrels Swift Observatory at 13:09:59 UT \citep{2021GCN.31202....1D}, the Gamma-ray Burst Monitor (GBM; \cite{2009ApJ...702..791M}) on board the Fermi Gamma-Ray Space Telescope at 13:09:59.651 UT \citep{GCN31210....1M} and High energy X-ray Telescope on board Insight-HXMT \citep{Xiao2022arXiv220502186X} at 13:09:59 UT on 2021 December 11. The burst is characterized by a spiky main emission phase lasting $\sim$13 s, and a longer, weaker extended emission  phase lasting $\sim$55 s \citep{Yang2022arXiv220412771Y}. The prompt emission is suggested to be produced  by the fast-cooling synchrotron emission \citep{Gompertz2022arXiv220505008G}.
The discovery of a kilonova associated with this GRB indicates clearly that the progenitor is a compact object merger \citep{Rastinejad2022arXiv220410864R}. The event fluence (10-1000 keV) of the prompt emission is
$(5.4 \pm 0.01)\times 10^{-4} {\rm \ erg \ cm^{-2}}$, making this GRB an exceptionally bright event. The host galaxy redshift of GRB 211211A  is $z =0.0763\pm0.0002$ (corresponding to a distance of $\approx$350 Mpc; \cite{Rastinejad2022arXiv220410864R}).   At 350 Mpc, GRB 211211A is one of the closest GRBs,  a bit further than GRB 170817A, which  is associated with the gravitational wave (GW)-detected binary neutron star (BNS) merger GW170817. For GRB 170817A, no GeV afterglow was detected by the Large Area Telescope (LAT) on timescales of minutes, hours, or days after the LIGO/Virgo detection \citep{Ajello2018ApJ...861...85A}.

As the angle from the Fermi-LAT boresight at the GBM trigger time of GRB 211211A is 106.5 degrees \citep{GCN31210....1M}, LAT cannot place constraints on the existence of high-energy ($E>100$ MeV) emission associated with the prompt GRB emission. We focus instead on constraining high-energy emission on the longer timescale. We analyze the late-time Fermi-LAT data when the GRB enters the field-of-view (FOV) of Fermi-LAT. We detect a transient source with a significance of $\rm TS_{max}\simeq 51$, corresponding to a detection significance over $6\sigma$. The result of the data analysis is shown in Section \ref{sect:data_analy} and the interpretation of the origin of this GeV afterglow is given in Section \ref{sect:inter} . In Section \ref{sect:sum} , we give a brief summary.

\section{Fermi-LAT data analysis}
\label{sect:data_analy}
At 13:09:59 UT (denoted as $\rm T_0$), the Swift/BAT triggered and located GRB 211211A \citep{2021GCN.31202....1D}.
The Fermi-LAT extended type data for the GRB 211211A was taken from the Fermi Science Support Center\footnote{\url{https://fermi.gsfc.nasa.gov}} from $\rm T_{0}-10 \ days$ to $\rm T_{0}+10 \ days$. 
Only the data within a $14\degr \times14\degr$ region of interest (ROI) centered on the position of GRB 211211A are considered for the analysis (initially centred on  Swift/X-ray Telescope (XRT)  position).
We select the observation time of GRB 211211A by LAT only when the angle from the Fermi-LAT boresight is less than $100^{\circ}$. The first observation starts at 395 s after Swift/BAT trigger, so we use the Source-class event selection, which has more stringent background rejection cuts than Transient-class events and is better suited for analyses of long time intervals and dimmer sources.
We also select the events with energies between 100 MeV and 10 GeV, with a maximum zenith angle of 100$\degr$ to reduce the contamination from the $\gamma$-ray Earth limb.

\begin{figure*}
\includegraphics[angle=0,scale=0.8]{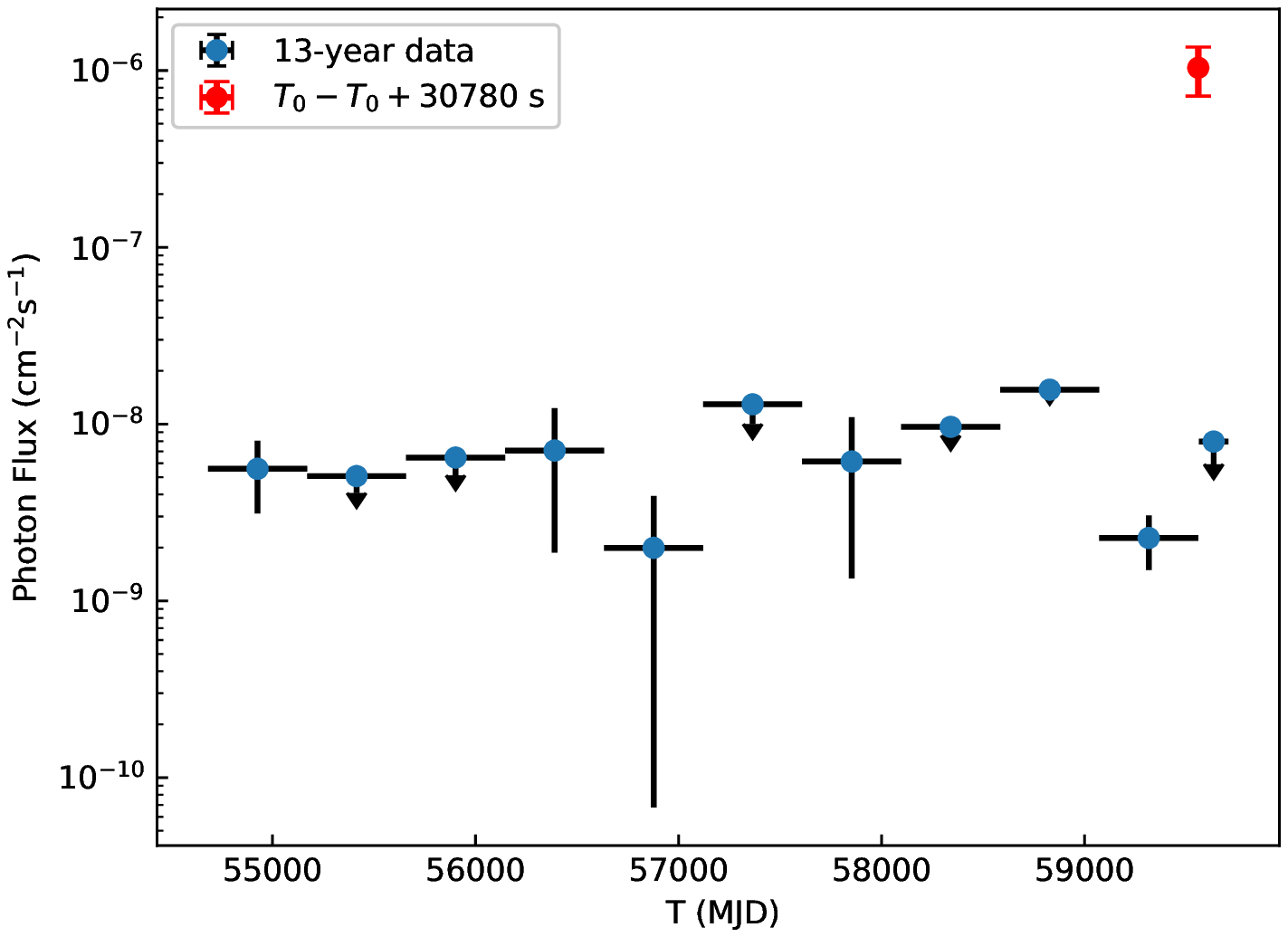}
\caption{Light curve of 4FGL J1410.4+2820 in 0.1-10 GeV (blue data) in comparison with the  GeV flux (red data) from GRB 211211A.
}
\label{LCAGN}
\end{figure*}

The instrument response functions (IRFs; P8R3\_SOURCE\_V3) is used.
The main background component consists of charged particles that are misclassified as gamma-rays. It is included in the analysis using the isotropic emission template (``iso\_P8R3\_SOURCE\_V3\_v1.txt'').
As the GRB is located at the high Galactic latitude, the contribution from the Galactic diffuse emissions is very small,  which is accounted for by using the diffuse Galactic interstellar emission template (IEM, gll\_iem\_v07.fits). The parameter of isotropic emission is left free and the one of IEM is fixed.
We note that one 4FGL source (4FGL J1410.4+2820) is close to  the GRB ($0.55\degr$ from Swift/XRT  position). 
We estimate its $\rm \sim 13.3$ years average flux before the GRB trigger to be $(1.59\pm0.41)\times 10^{-9} \ \rm photons \ cm^{-2}\ s^{-1}$ (the light curve of 4FGL J1410.4+2820 is shown in Figure \ref{LCAGN}). This flux implies that this source is not bright enough to be considered as the background source in the model.

\begin{figure*}
\includegraphics[angle=0,scale=0.6]{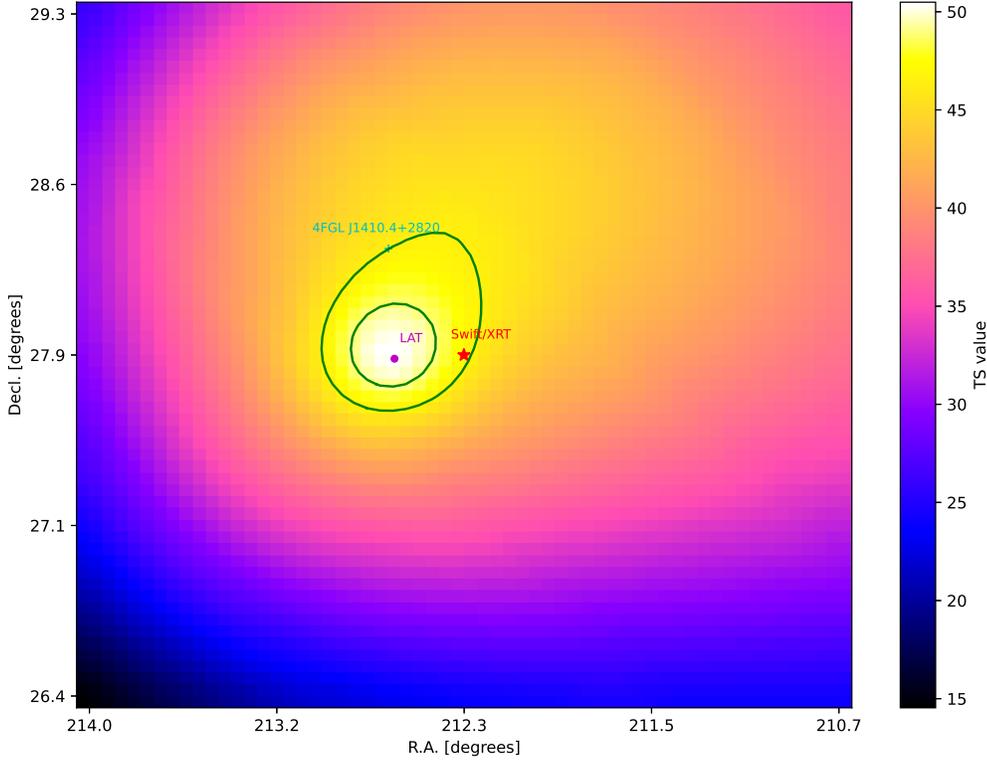}
\caption{$3\degr \times 3\degr$ TS map of the gamma-ray emission in 0.1--10 GeV measured by Fermi-LAT around GRB 211211A in 395--30780 s after the BAT trigger.
The cyan cross represents 4FGL J1410.4+2820, which is suggested to be associated with a BL Lacertae RX J1410.4+2821 by Fermi-LAT Collaboration \citep{2020ApJS..247...33A}.
The magenta point indicates the best localization of GRB 211211A. The two green lines represent the localization contours of GRB 211211A at 68\% and 90\% confidence levels, respectively.
The red star represents the localization of GRB 211211A  by Swift/XRT \citep{2021GCN.31202....1D}.
}
\label{GeVmap}
\end{figure*}

The data analysis was performed using the publicly available software fermitools  (ver. 2.0.8) with the unbinned likelihood analysis method. Assuming a power-law spectrum of the burst, we obtained the best-fit { Fermi-LAT} position of GRB 211211A in 395--30,780 s with the tool gtfindsrc: (212.60$\degr$, 27.87$\degr$) with a circular error of $0.20\degr$ (statistical only).
The maximum likelihood test statistic (TS) is used to estimate the significance of the GRB, which is defined by TS$= 2 (\ln\mathcal{L}_{1}-\ln\mathcal{L}_{0})$, where $\mathcal{L}_{1}$ and $\mathcal{L}_{0}$ are maximum likelihood values for the background with the GRB and without the GRB (null hypothesis).
Using the gttsmap tool, we evaluate the TS map in the vicinity of the GRB, which is shown in Figure \ref{GeVmap}.
The maximum value, $\rm TS_{max} = 50.82$, is found at the location of $\rm R.A. = 212.61\degr$ and $\rm Decl. = 27.91\degr$ (J2000), consistent with the result of gtfindsrc. This $\rm TS_{max}$ value corresponds to a detection
significance of $6.2\sigma$ or $6.7\sigma$ (one-sided) if the $\rm TS_{max}$ distribution follows $(1/2)\chi^{2}_{4}$ or $(1/2)\chi^{2}_{2}$, respectively
{\footnote {As interpreted in the first LAT GRB catalogue \citep{2013ApJS..209...11A}, the model for a GRB analysis usually has 4 degrees of freedom, i.e., the two coordinates (e.g. (R.A., Decl.)) of the GRB  and  two spectral parameters. However, when an external position is used in the analysis (for example, the Swift/XRT initial position here), only 2 degrees of freedom are left.}}.
We compute the error contours of the source localization using the method suggested by \cite{2021NatAs...5..385F}, and the iso-contours containing localization probabilities of 68\% and 90\% are shown as green lines in Figure \ref{GeVmap}.
We find that the GRB position detected by Swift is inside the region of the localization contours of LAT at the 90\% confidence level. 

As part of our analysis, we use the gtsrcprob tool to estimate the probability that each photon detected by the LAT is associated with the GRB.
The list of events associated with the GRB with a probability higher than 80\% is shown in Table \ref{tab:prob}, among which six photons have the probability higher than 90\%. 
The first $\gamma$-ray photon with the probability exceeding 90\% arrives at $\rm T_{0}+ 6438.83\ s$, with an energy of 206.91 MeV, and the highest-energy  photon is a 1740.45 MeV photon arriving at $\rm T_{0}+ 12967.39\ s$. 

\begin{figure*}
\includegraphics[angle=0,scale=0.6]{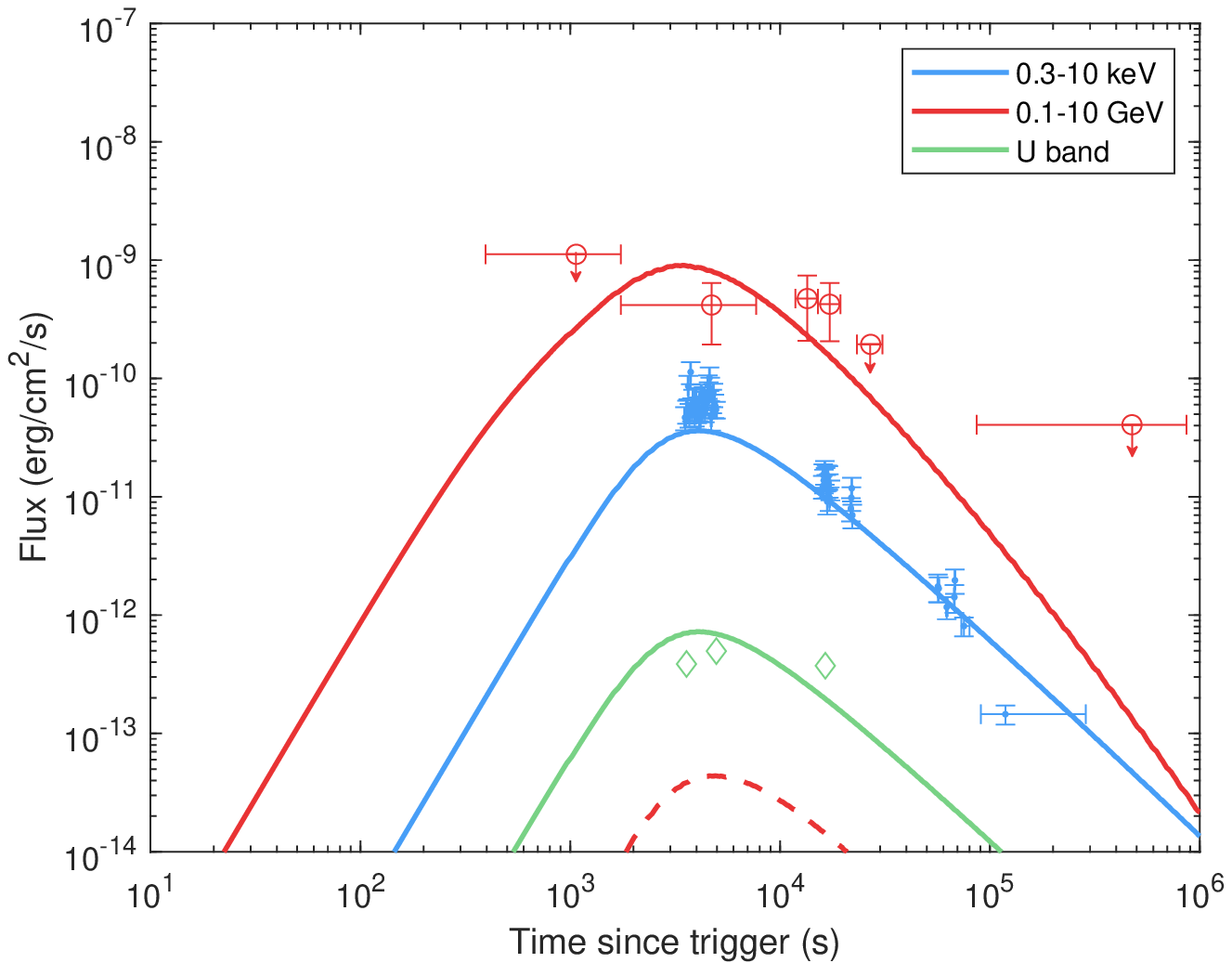}
\includegraphics[angle=0,scale=0.6]{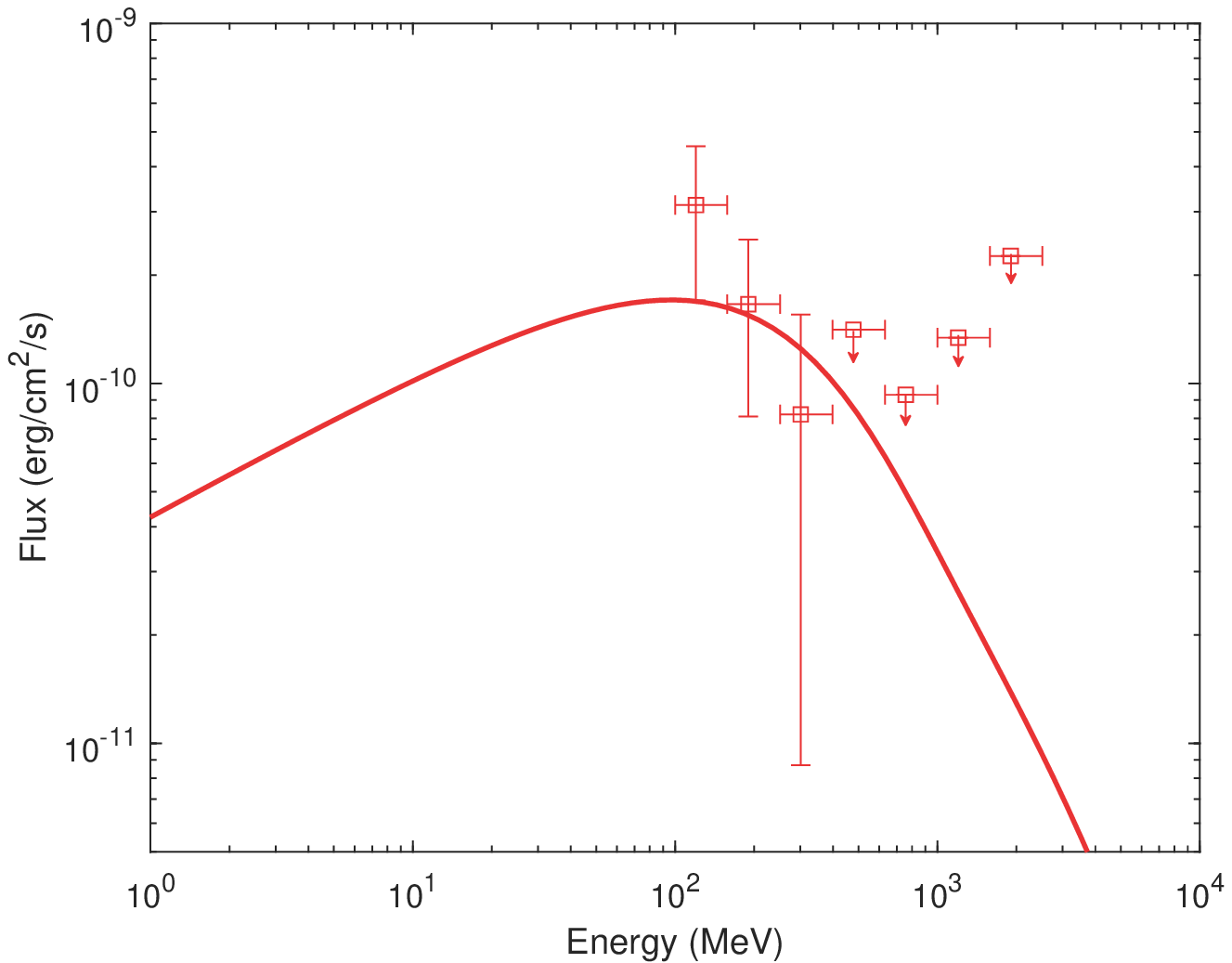}
\caption{Left panel: light curves of the GeV emission of GRB 211211A measured by Fermi-LAT and the modeling of the multi-wavelength afterglow light curves. The red, blue and green data points represent the GeV, X-ray and early optical flux of GRB 211211A, respectively. X-ray data are downloaded from the UK Swift Science Data Centre (UKSSDC; \cite{2007A&A...469..379E,2009MNRAS.397.1177E}). The optical data are obtained from \cite{Rastinejad2022arXiv220410864R}.
The solid  lines represent the synchrotron emission at GeV, X-ray and optical bands, while the dashed line represents the SSC component at the GeV band in our modeling. 
Right panel:  the measured spectrum and modeling of the GeV emission of GRB 211211A  during 395--30780 s. 
The parameters used in the modeling are $E_{\rm k, iso}=1\times 10^{53}{\rm \ erg}$, $\Gamma_0=100$, $n=10^{-4}{\rm \ cm^{-3}}$, $p=2.2$, $\varepsilon_{\rm e}=0.1$, $\varepsilon_{\rm B}=6\times 10^{-5}$, $\theta_{\rm j}=1.0^{\circ}$ and $\eta_{\rm acc}=0.01$ (see the text for more details).
}
\label{lcsed}
\end{figure*}

The averaged flux is $(3.23\pm0.86)\times 10^{-10}\ {\rm erg \ cm^{-2} \ s^{-1}}$ with a photon index $\Gamma_{\rm LAT}=-3.30\pm0.45$ in 395--30,780 s. Assuming that the GRB occurs at a distance of 350 Mpc, the measured value of the flux corresponds to a luminosity of $(4.74\pm1.26)\times 10^{45}\ {\rm erg \ s^{-1}}$.
Figure \ref{lcsed} shows the temporal behavior and the spectrum of the  GeV emission from GRB 211211A.

\section{Interpretation of the GeV emission}
\label{sect:inter}
Extended high-energy ($>100$ MeV) gamma-ray emission that lasts much longer than the prompt sub-MeV
emission has been detected from a large number of GRBs by Fermi–LAT.  Before Fermi, extended GeV emission has been detected from GRB 940217 by EGRET \citep{Hurley1994Natur.372..652H}, with one 18\,GeV photon arriving at about 5000\,s after the burst.   A plausible scenario for the extended high-energy emission is that it is the afterglow synchrotron emission produced by electrons
accelerated in the forward shocks \citep{Kumar2010MNRAS.409..226K,Ghirlanda2010A&A...510L7G,Wang2010ApJ...712.1232W}. In particular,  high-energy ($>100$ MeV) emission is
detected up to $\sim100$ s in the short GRB 090510 \citep{Abdo2009}. The multiwavelength (0.1-10 GeV, X-ray, and optical)
emission of GRB 090510 can be  explained via synchrotron emission from an adiabatic forward shock propagating into
a homogeneous ambient medium \citep{He2011ApJ...733...22H}.  However, since the maximum synchrotron photon energy for accelerated electrons under the most favorable condition
(i.e., the Bohm acceleration) is about 50 MeV in the shock rest
frame and the bulk Lorentz factor of the external shock decreases with time, it is a challenge to explain $>10$ GeV photons detected from GRB afterglows  \citep{Duran2011MNRAS.412..522B,Piran2010ApJ...718L..63P,Sagi2012ApJ...749...80S}. 
In fact, recent detection of sub-TeV emission from a few GRBs has been interpreted as arising from the inverse-Compton (IC) emission of the afterglow \citep{MAGICGRB190114TeV,Derishev2021ApJ...923..135D,Wang2019ApJ...884..117W}.

The maximum energy of the photons detected from GRB 211211A is only 1.7 GeV, so it is reasonable to consider the afterglow synchrotron emission scenario. 
Below we study the possibility of the forward shock  emitting the GeV afterglow emission.  
We perform modeling of the Fermi-LAT data for GRB 211211A using a  numerical code developed in our previous work \citep{Liu2013ApJ...773L..20L}. According to the standard afterglow model \citep{Sari1998ApJ...497L..17S}, the light curve for a given observed frequency ($\nu$) could be calculated as
$F(t,\nu)=F(t,\nu,E_{\rm k, iso},n,p,\varepsilon_{\rm e},\varepsilon_{\rm B},\Gamma_{0}, \theta_{\rm j})$. Here $E_{\rm k, iso}$ is the isotropic kinetic energy of the GRB outflow, $n$ is the particle number density of the ambient medium, $p$ is the electron spectral index, $\varepsilon_{\rm e}$ and $\varepsilon_{\rm B}$ are the equipartition factors for the energy in electrons and magnetic field in the shock,  $\Gamma_0$ is the initial Lorentz factor of the outflow, and $\theta_{\rm j}$ is the half-opening angle of the jet. In this code, the electrons that produce synchrotron high-energy emission also undergo IC loss and the Klein–Nishina (KN) effect has been taken into account \citep{Wang2010ApJ...712.1232W}. We find that the model can reproduce the light curves of  GeV and X-ray afterglows, as well as the optical afterglow at early time when the kilonova emission is subdominant \citep{Rastinejad2022arXiv220410864R},   for the following parameter values: $E_{\rm k,iso}=1\times 10^{53}{\rm \ erg}$, $\Gamma_0=100$, $n=10^{-4}{\rm \ cm^{-3}}$, $p=2.2$, $\varepsilon_{\rm e}=0.1$, $\varepsilon_{\rm B}=6\times 10^{-5}$ and $\theta_{\rm j}=1.0^{\circ}$ (see Figure \ref{lcsed}). The measured photon index of the X-ray afterglow by Swift/XRT ($\Gamma_{\rm X}=-1.5^{+1.2}_{-0.06}$; \cite{2021GCN.31212....1O})  is also consistent with  $p=2.2$, as the X-ray frequency locates in the regime $\nu_m<\nu_{\rm X}<\nu_c$, where $\nu_m$ and $\nu_c$ are, respectively, the frequencies corresponding to the injection break and cooling break \citep{Sari1998ApJ...497L..17S}. We note that the observed X-ray flux at 3-5 ks exceeds the model flux to some extent, which could indicate that the early X-ray emission  may have some contribution from an extra component other than the afterglow, such as  the central engine activity, as have been seen in some GRB X-ray afterglows \citep{2007ApJ...665..599T}. 

To explain the long-duration of the GeV emission, a long deceleration time for the external shock is needed. This implies a low density of $n\simeq 10^{-4}{\rm \ cm^{-3}}$. Such a low density of the ambient medium is not surprising,  since the GRB lies outside of the optical disk of the host galaxy, consistent with the  compact stellar merger scenario. In addition, Quasi-Periodic Oscillations (QPO) with frequency  $\simeq 22$ Hz are found throughout the precursor of GRB 211211A \citep{Xiao2022arXiv220502186X}, which indicates most likely that a magnetar participated in the merger. The pulsar wind from the magnetar may have created a cavity around the pulsar \citep{Holcomb2014ApJ...790L...3H}. 

The spectrum measured by Fermi-LAT appears softer than the predicted synchrotron spectrum $F_{\rm GeV}\sim \nu^{-p/2}$. One solution to this problem is assuming a maximum cutoff energy for the synchrotron emission, which is related to the maximum energy of shock-accelerated electrons. By equating the synchrotron cooling time with the acceleration time, we get the maximum Lorentz factor of the accelerated electrons, $\gamma_{M}=\sqrt{(6\pi e \eta_{acc})/(\sigma_{\rm T} (1+Y)B)}$, where $\eta_{acc}$ is the acceleration efficiency, $\sigma_{\rm T}$ is the Thompson scattering cross section, $Y$ is the Compton parameter for IC emission and $B$ is the magnetic field.  We find that assuming $\eta_{acc}=0.01$, the GeV spectrum of GRB 211211A can be reproduced  by the model. This implies that the Bohm acceleration approximation breaks down at such high energies, possibly due to the small-scale nature of the microturbulent magnetic field behind the shock \citep{Wang2013ApJ...771L..33W}.

\section{Summary}
\label{sect:sum}
We reported the detection of a GeV afterglow from GRB 211211A, which is a long-duration GRB, but results from a compact stellar merger. The GeV emission continues up to about 20,000\,s after the burst. The duration is the longest one compared to the GeV afterglows of other short-duration GRBs (see the Extended Data Fig. 7 in \cite{2021NatAs...5..385F}). However,  it is quite similar to the long-duration GRB 940217 \citep{Hurley1994Natur.372..652H}, which has one 18\,GeV photon at about 5000\,s after the burst. It has been suggested that the late GeV emission of GRB 940217 may be produced by the synchrotron-self Compton (SSC) emission of the afterglow \citep{Dermer2000ApJ...537..785D,ZhangApJ...559..110Z}. However, for GRB 211211A, it is hard to reproduce the GeV peak at 20,000\,s with SSC emission given the constraint from the X-ray and optical afterglows. In addition, since GRB 211211A occurs at the position outside the galaxy disk, the density of the circumburst medium is expected to be low, which leads to a subdominant contribution to the GeV emission by the SSC component. Instead, we find that the  GeV   emission of GRB 211211A can be interpreted as arising from the afterglow synchrotron emission. The soft spectrum of the GeV emission could arise from a limited maximum synchrotron energy of only a few hundreds of MeV at $\sim 20,000$\,s. The long duration of the GeV emission can be interpreted as the long deceleration time due to a low-density circumburst environment, which agrees well with the density environment expected for compact stellar mergers.

\acknowledgments

The work is supported by the NSFC
Grants No.12121003 and No. U2031105, 
the National Key R$\&$D program of China under Grant
No. 2018YFA0404203, and China Manned Spaced Project
(CMS-CSST-2021-B11).

While we were preparing the final submission, we became
aware of the work by \cite{2022arXiv220508566M}, which also reports the
detection of GeV emission from GRB 211211A. The two
works are independent of each other.

\bibliography{reference}{}
\bibliographystyle{aasjournal}

\begin{table}[ht!]
\caption{List of the selected gamma-ray events with a probability of association $>80\%$ in 395-30780 s. }
\begin{center}
    \begin{tabular}{lcccc}
        \hline
        Time since $T_{0}$ (s) & Energy(MeV) &R.A.(deg) & Decl.(deg) & Probability($\%$) \\ \hline
1165.89 & 400.05 & 213.28 & 27.25 & 86.80 \\
6238.33 & 142.38 & 212.64 & 29.01 & 84.18 \\
6438.83 & 206.91 & 212.67 & 27.82 & 90.42 \\
6648.08 & 187.58 & 212.11 & 28.91 & 90.91 \\
12494.06 & 163.96 & 212.57 & 29.20 & 89.95 \\
12967.39 & 1740.45 & 212.63 & 27.85 & 98.91 \\
13054.08 & 102.64 & 211.48 & 28.20 & 94.41 \\
17410.06 & 113.83 & 211.61 & 27.22 & 92.45 \\
17861.10 & 286.10 & 213.34 & 28.65 & 88.36 \\
18128.17 & 231.49 & 212.80 & 28.36 & 94.83 \\
24487.04 & 104.01 & 214.33 & 29.19 & 88.31 \\
28335.50 & 274.59 & 212.37 & 26.89 & 88.35 \\
         \hline
    \end{tabular}
    \end{center}
\label{tab:prob}
\end{table}{}

\end{document}